\documentclass[a4paper,11pt]{article}
\pdfoutput=1 

\usepackage{jheppub} 

\usepackage[T1]{fontenc} 
\usepackage{amsmath,amssymb}

\usepackage{graphicx}
\usepackage{epstopdf}
\DeclareGraphicsExtensions{.eps}
\usepackage{comment}

\newcommand{\ba}{\begin{eqnarray}}
\newcommand{\ea}{\end{eqnarray}}
\newcommand{\nn}{\nonumber}

\newcommand{\cO}{{\mathcal{O}}}

\newcommand{\be}{\begin{equation}}
\newcommand{\ee}{\end{equation}}

\def\a{\alpha}
\def\b{\beta}

\def\d{\delta}
\def\e{\epsilon}

\def\t{\tau}

\def\y{\eta}
\def\z{\zeta}

\allowdisplaybreaks[1]
\title{\boldmath Classical Virasoro irregular conformal block II}

 \author{Chaiho Rim}
 \author{and Hong Zhang}
 \affiliation{Department of Physics and Center for Quantum Spacetime (CQUeST),\\
Sogang University, Seoul 121-742, Korea}

\emailAdd{rimpine@sogang.ac.kr}
\emailAdd{kilar@sogang.ac.kr}

\abstract{We present a new systematic way to 
evaluate the classical limit 
of the Virasoro  irregular conformal block for arbitrary rank n
based on the irregular partition function. 
In addition, we  prove  that 
the classical irregular conformal block has the exponential form
as  suggested by  A. Zamolodchikov and Al. Zamolodchikov 
for the regular case. 
We provide an explicit calculation for  the rank 2 case in detail. }

\begin{document} 
\maketitle
\flushbottom

\setcounter{footnote}{0}


\section{Introduction} 

In our recent work \cite{RZ} 
which will be called I in the following,
we studied the classical limit (NS limit) \cite{NS} 
of Virasoro irregular conformal block  (ICB) using the irregular matrix model (IMM)
 \cite{CRZ} .
IMM is a $\beta$-deformed  one matrix model with a logarithmic  
as well as a finite number of  inverse power potentials. The finite number $n$
of the inverse powers   is called the 
rank of irregular model. 
It is demonstrated that the classical  ICB  of rank $n$ can be obtained 
by using the generalized Mathieu equation.
This equation is equivalent to the loop equation of IMM 
and is solved   on a unit circle with the Floquet exponent for the rank 1.
However, this method is not easy to generalize to the case with arbitrary rank n.
In this paper, we present a new systematic way to find ICB 
based on the loop equation. 
  
In section \ref{proof}  we present that
the classical  irregular conformal block  ${\cal F}_\Delta^{(m:n)} $, 
the inner product between irregular modules of rank $m$ and $n$
has the  exponential form:
\begin{equation}
{\cal F}_\Delta^{(m:n)} 
\;\stackrel{g_s \to 0}{\sim}\;
\exp\left\lbrace \frac{1}{g_s^2} f_{\delta}\right\rbrace,   
\label{eq_e-conjecture}
\end{equation}
where $g_s \to 0$ corresponds to the  classical limit 
and $f_{\delta} $ is  finite in this limit.
This is suggested in \cite{ZZ, LLNZ_201309}  for the regular conformal block
and extended to the ICB in  \cite{ PP_201407}.
We demonstrate this exponential behavior  
for the regular and irregular case 
using the conformal property of the loop equation.
In section \ref{partition} we present a new systematic way to analyze the 
classical limit of IMM and give a non-trivial  example for the rank 2 partition function $Z_{(0:2)}$. 
Further in section \ref{con-block} we evaluate the partition function $Z_{(m:n)}$ based on the same method, which is sufficient to construct ICB.
Section \ref{conclusion} is the conclusion and the appendix deals with the normalization
of the partition function and technical details.

\section{Classical form of conformal block} 
\label{proof}
\subsection{Setup of the formalism}
Regular matrix model  \cite{DV_200909, {Itoyama}} 
related with the regular conformal block is defined as
the $\beta$-deformed Penner-type matrix model
\be
Z_{\beta}   =     \int \left( \prod_{I=1}^N d \lambda_I \right)
 \prod_{I < J} (\lambda_I - \lambda_J)^{2\b} 
            e^{ \frac{\sqrt{\beta}}{g}  \sum_I V(\lambda_I)}\,.
\label{eq_beta-matrix}
 \ee
$V(\lambda)$  is the Penner-type potential 
\be
  V(z )  =     \sum_{a=0}^{K} \hat \a_a \log (z - z_a)\,.
\ee
This potential is obtained from the correlation of 
$K+2 $  primary vertex operators (lying at $0, z_1, \cdots, z_K, \infty$)
and screening operators (lying at $z$).
$\beta$ is related with the Virasoro screening charge $b= i \sqrt{\beta}$
and  the  Virasoro charge  $ \alpha$ 
of the primary operator is rescaled as $ \hat  \alpha =g_s \alpha $.
We introduce the small expansion parameter $g$ 
which is related with  $g = ig_s/2$ 
so that $\sqrt{\beta} /g = - 2 b /g_s$. 

Classical limit is obtained as $g_s\to 0$ 
so that $\hat \alpha_a$ is finite. On the other hand,
$b$ and $g_s$ are related with the $\Omega$ deformation 
parameter $\epsilon_1= g_s b$ and $\epsilon_2= g_s/b$
of the Nekrasov partition function \cite{Nekrasov1,Nekrasov2, NO}
according to AGT conjecture \cite{AGT}.
Therefore, classical limit is achieved either  
$\epsilon_2 \to 0$ but $\epsilon_1$ finite
which is the  Nekrasov-Shatashvili (NS) limit or its dual
($\epsilon_1 \to 0$ but $\epsilon_2$ finite). 
The two pictures are equivalent since Liouville theory has 
$b \to 1/b$ duality.

IMM is obtained in  \cite{TR_201207,CR} from the colliding limit \cite{{EM_2009}, GT_2012}.
IMM has the same form of   \eqref{eq_beta-matrix} 
but the potential is different,
\be 
 V   (z )
=   \hat c_0 \log z - \sum_{k=1 }^{n}   \left( \frac {{ \hat c_k} } {k z^{k}} \right)   
+ \sum_{\ell=1}^{m}   \left( \frac { \hat c_{-\ell}~ z^{\ell}}   {\ell}  \right )  \,,
\label{irregular_potential}
\ee
where  $m+n=K$ and $(m+1)$ primary operators are put to $\infty$, 
and $(n+1)$ operators to $0$.
The coefficients  $\hat c_k$ and $\hat c_{-\ell}$ 
are given in terms of the moments;
$  \hat c_k= \sum_{r=1}^{n} \hat \a_r(z_r)^k  $
with $k\ge 0$
and  $  \hat c_{-\ell}= -\sum_{a=1}^{m} \hat \a_a(z_a)^{-\ell}   $ 
with $\ell>0$.  
The irregular partition function  with the   potential \eqref{irregular_potential}
will be denoted as $Z_{(m;n)}(\hat c_0; \{ \hat c_k\}, \{ \hat c_{-\ell} \})$ in the following.

The matrix model  (regular or irregular) has the loop equation 
which presents the symmetric property 
\be
4 W(z)^2 + 4 V'(z) W(z) + 2  g_s Q W'(z) -  g_s^2 W(z,z) =f(z) \,,
\label{eq_loop}
\ee
where $Q=b+1/b$ is the background charge. 
$ W(z)$ and $ W(z,z)$ are the one and two point resolvents,
defined as  
$ W(z) = g \sqrt{\beta}   \left\langle\sum_I  \frac{1} {z - \lambda_I} \right\rangle_{\!\!\!conn}$
and
$ W(z,w) = \beta   \left\langle\sum_I 
 \frac{1} {(z - \lambda_I)(w - \lambda_I) } \right\rangle_{\!\!\!conn}$,
respectively. 
The bracket  $\left\langle O \cdots \right \rangle_{\!\!\!conn} $denotes the  
connected part of the  expectation value 
with respect to the matrix model \eqref{eq_beta-matrix}. 
$f(z)$ is the  expectation value determined by  the potential $V(z)$,
$
 f(z)    =   4g \sqrt{\beta} \left\langle
 \sum_I \frac{V'(z)- V'(\lambda_I)}{z - \lambda_I} 
\right\rangle_{\!\!\!conn}\,.
$

At the classical limit, the resolvents defined above remain  finite  \cite{MMM}.
Therefore, the loop equation \eqref{eq_loop} is simplified as  
\be
       x(z)^2 +  \epsilon x'(z)  + U(z) =0 \,,
\label{eq_loop-i}  
\ee 
where $x(z) = 2 W(z) + V'(z)$ and 
$
U (z)
=  - \Big (V'(z) \Big) ^2   -{\epsilon}  V''(z)  - f(z) 
$.
$\epsilon= g_s Q$ is a finite parameter at the classical/NS limit. It should be noted 
that \eqref{eq_loop-i}  is manifestly invariant for both limits (NS and its dual).
This loop equation turns  into a second order differential equation
(Shr\"odinger-like equation)
if one defines  
 $  \Psi (z) 
     =     \exp \left( \frac{1}{\epsilon} \int^{z} x(z') dz' \right)$:
    \be
   \left ( \epsilon^2 \frac{\partial^2}{\partial z^2}   + U(z) \right)  \Psi(z) =0\,.
\label{schrodinger}
    \ee

On the other hand, one may conveniently
investigate the conformal block using a degenerate primary operator.
In  \cite{RZ, BMT_201104}, an  expectation value  
$
P(z) \equiv    \Big\langle  \prod_I  ( z-\lambda_I ) \Big\rangle  
$
is introduced in relation with the  degenerate operator.  
It should be noted that  $P(z)$ is  a polynomial of degree $N$,
 \ba
  P(z) = P_0+ P_1 z +P_2 z^2+ \dots+P_{N-1}z^{N-1}+ P_N z^{N}.
 \ea
where $N$ is the number of integration variables 
in \eqref{eq_beta-matrix} 
and $P_N$ is normalized to be 1. 

The wave function  $ \Psi(z)$ in  \eqref{schrodinger}
is closely related with $P(z)$.
This can be seen if one notes that at the classical limit, one has  \cite{RZ} 
\be 
\log \left( \frac { P (z)}{ P (z_0)} \right) 
 =  \frac2 {\epsilon }  
\int_{z_0}^z      dz' ~W(z' )  \,.
\ee
Taking derivatives one has $ W(z)=\frac{\epsilon}{2} \left( \log { P (z)} \right) '
 =\frac{\epsilon}{2}\frac { P' (z)}{ P (z)} $ and therefore, 
the wave-function $ \Psi (z)$ is given as  
\be 
 \Psi (z)
     =    P (z) \exp \left( \frac{1}{\epsilon} \int^{z} V'(z') dz' \right) \,.
\ee 
This shows that the polynomial function satisfies the second order differential equation 
\begin{align}
\label{maineq}
{\epsilon}^2  P''(z)   +  2{\epsilon} V'(z) P'(z)  = f(z) P(z)\,,
\end{align} 
which can be check  from \eqref{eq_loop-i} or equivalently from  \eqref{schrodinger}.
We will use this equation to investigate the partition function  and conformal block. 

\subsection{Classical irregular conformal block} 
Let us investigate the exponential behavior of the classical conformal block.
Note that \eqref{eq_e-conjecture} is equivalent to that
\be 
\lim\limits_{g_s \to 0} g_s^2 \log {\cal F}_\Delta^{(m:n)} 
\to
finite\,.
\ee
In this section we concentrate on the case of ICB.
Regular conformal block is commented in sec \ref{regular}.

The explicit form of ICB is given in tems of IMM $Z_{(m;n)}$  \cite{CRZ}: 
\be
{\cal F}_\Delta^{(m:n)}  (\{\hat c_{-\ell }  :  \hat c_k \})= 
\frac{e^{\zeta_{(m:n)}}   Z_{(m:n)}  (\hat c_0; \{\hat c_k\}, \{\hat c_{-\ell}\})} 
{Z_{(0:n)}(\hat c_0;  \{ \hat c_k \})  Z_{(0:m)} ( \hat c_\infty; \{ \hat c_{-\ell }\}) }\,,
\label{cb}
\ee 
where $c_0$ is fixed by the neutrality condition $c_0 + c_{\infty} + Nb =Q$
with $N$ the number of inserted screening operators.
ICB has an extra factor $e^{\zeta_{(m:n)}}$
which comes from  the limiting procedure 
$z_a \to \infty$ and $z_b \to 0$. 
Explicitly $\zeta_{(m:n)}=\hat \zeta_{(m:n)}/g_s^2$
where $\hat \zeta_{(m:n)}=\sum_k^{{\rm min}(m,n)} 2 \hat c_k \hat c_{-k}/k$.

One can confirm the exponential behavior \eqref{eq_e-conjecture} 
using the expression of ICB.
We need to confirm the classical behavior  
\be 
\lim\limits_{g_s \to 0} g_s^2\left\lbrace {\zeta_{(m:n)}}+  \log Z_{(m:n)} 
-  \log Z_{(0:n)} -  \log Z_{(0:m)} \right\rbrace
\to  finite \,.
\ee
It is easy to show that the first term is finite since it is given as 
 \be  \lim\limits_{g_s \to 0}~~ g_s^2 ~ \zeta_{(m:n)}  
= \hat \zeta_{(m:n)} \,.
\ee

The contribution of $Z_{(0:n)} $ 
can be  evaluated using  $f(z)$.
Note that $f(z)$ 
has a finite number of inverse powers of $z$:
$f(z)= \sum_{k=0}^{n-1}  {d_k } {z^{-(k+2)}}$. 
Therefore, if one  expands  Eq. \eqref{maineq}   in powers of $z$, 
one finds the equation has the terms
running  from $z^{N-2}$ to $z^{-n-1}$. 
This  provides  $N+n$ number of equations.
Since there are $N+n$ 
unknown variables: $P_0$, $P_1,$ $\dots,$ $P_{N-1}$ 
and $d_0$, $d_1,$ $\dots,$ $d_{n-1}$,
one can solve the equations 
to find $d_k$ as a  function of $\hat c_k$'s,
which are finite at the classical limit $g_s \to 0$.
Once the solution of $d_k$ is found, one can
find the partition function $Z_{(0:n)} $ 
using the differential equation   \cite{TR_201207}
\be
\label{fd}
-g_s^2~  v_k ( \log   {Z_{(0:n)}}) = d_k  ~~~{\rm for}~0\le k \le n-1\,,
\ee
where  $v_k$ 
is the differential operator related with the Virasoro generator representation:
\be
 v_{k\ge0} =  \sum_{\ell>0}  \ell~\hat c_{\ell+k }  \frac{ \partial}{\partial \hat c_\ell} \,.
\label{v_k}
\ee 
Here we use the  convention  $\hat c_\ell=0$  when $c_\ell$ does not belong to 
$\{\hat c_0, \cdots, \hat c_n\}$. 
Once $d_k$ is known, one may rearrange \eqref{fd} to put 
\be
\label{ffd}
 -g_s^2  \frac{ \partial}{\partial \hat c_\ell} \log   {Z_{(0:n)}}=  F_\ell(\{\hat c_k\} )~~~~
{\rm for}~ 1 \le \ell \le n\,.
\ee 
Since $Z_{(0:n)} $ only depends on $\hat c_0$, $\hat c_1,$ $\dots,$ $\hat c_{n}$, \eqref{ffd} 
is sufficient to determine ${Z_{(0:n)}}$ completely, 
up to the normalization  factor $N_{(0:n)}$ 
which is independent of   $\hat c_{\ell>0}$:
\ba \label{0nh}
 -g_s^2   \log   \left( \frac {Z_{(0:n)}} {N_{(0:n)}}  \right) 
=H_{(0:n)}(\hat c_0, \{\hat c_k\} )  \,,
\ea
with finite $H_{(0:n)}$ at the classical/NS limit.

In a similar way, one has $f(z)= \sum_{k={-m}}^{n-1}  {d_k } {z^{-(k+2)}}$ for $Z_{(m:n)}$.
By identifying each coefficient of $z^\ell$ in \eqref{maineq}, there are $N+m+n$ equations, running  from $z^{N+m-2}$ to $z^{-n-1}$. 
The number of unknown variables are also $N+m+n$ 
: $P_0$, $P_1,$ $\dots,$ $P_{N-1}$ 
and $d_{-m}$, $\dots,$$d_{-1},$  $d_0$, $d_1,$ $\dots,$ $d_{n-1}$. 
Thus solutions of $d_k$ exist  as functions of $\hat c_k$.
Furthermore this coefficients allows 
to find $Z_{(m:n)}$ through the differential equation  \cite{CR}
\begin{align}
 -g_s^2~ v_k ( \log   {Z_{(m:n)}}) & =  d_k  \qquad\qquad~~~~~~~{\rm for}~0 \le k \le n-1\,,
\nn\\
 -g_s^2~ u_k ( \log   {Z_{(m:n)}}) & =  d_{-k}  -2 \epsilon N \hat c_{-k} ~~~~{\rm for}~1 \le k <m-1\,,
\label{uv_k}
\end{align} 
where $u_k$ is the differential operator corresponding to 
$\hat c_{-\ell}$ 
\be
 u_{k>0} =  \sum_{\ell >0}  \ell~\hat c_{-\ell -k }  \frac{ \partial}{\partial \hat c_{-\ell}} \,.
\label{u_k}
\ee
The solution is found similar to \eqref{0nh}, 
\ba 
 -g_s^2   \log   \left( \frac {Z_{(m:n)}} {N_{(m:n)}}  \right) 
=H_{(m:n)}(\hat c_0, \{\hat c_k\},\{\hat c_{-\ell} \})  \,,
\label{mnh}
\ea
with finite $H_{(m:n)}$ at the classical/NS limit.

In addition, the conformal block \eqref{cb} is defined as 1 if $ \hat c_k = \hat  c_{-\ell}=0$ 
for $k,\ell>0$. Therefore, the conformal block is independent of the normalization. 
(In appendix \ref{Normalization} we present how the normalization behaves at the classical/NS limit). 
Collecting all the terms, one has the classical ICB in the form of
\be \label{logf}
\lim\limits_{g_s \to 0} g_s^2 \log {\cal F}_\Delta^{(m:n)} 
=\hat \zeta_{(m:n)} - H_{(m:n)} + H_{(0:n)}+H_{(0:m)}
\ee
which is finite and thus, \eqref{eq_e-conjecture} is proved.
In the following sections, we present explicit form of $H_{(0:n)}$ and $H_{(m:n)}$
which is indeed finite.

\subsection{Classical regular conformal block} 
\label{regular}
We may demonstrate  that
\eqref{eq_e-conjecture} holds for the classical regular conformal block too. 
For the regular case, we still have \eqref{maineq}, but with 
$
V'(z) =  \sum_{a=0}^{K} \frac { \hat \a_a } {z-z_a}
$ and 
\ba
f(z) =      \sum_{a=0}^{K} \frac {d_a } {z-z_a}, \qquad \qquad d_a=-g_s^2\frac{\partial \log Z_{\beta}}{\partial z_a}\,.
\ea

We start with the equation of $P(z) $ in \eqref{maineq}. 
If one   takes the residue of \eqref{maineq} around each $z_a$,
one obtains $K+1$ equations for $a=0, 1, \dots, K$:
\ba
 2{\epsilon} \hat \a_a P'(z_a)  =d_a P(z_a)\,.
\ea
Equivalently,
\ba
 2{\epsilon} \hat \a_a \frac{\partial \log P(z_a)}{\partial z_a}  =-g_s^2\frac{\partial \log Z_{\beta}}{\partial z_a}\,.
\ea
Thus, one has 
\ba
 Z_{\beta} =N_{\beta} \prod_{a=0}^n{ P(z_a)}^{-2{\epsilon} \hat \a_a/g_s^2} \,,
\ea
where $N_{\beta} $ is the normalization factor independent on $z_a$ and is discussed in the appendix.
One may also normalize the conformal block so that the $z_a$-independent factor as 1.

$P(z)$ is given as the solution of \eqref{maineq}.
If the solution exists, the solution should be finite. There is no singularity forbidding $z \to z_a$ 
since $P(z)$ is the polynomial with degree $N$. 
Thus, one may conclude  for the regular conformal block
$ Z_{\beta} \;\stackrel{g_s \to 0}{\sim}\;
\exp\left\lbrace   \z/ {g_s^2} \right\rbrace
$ 
where 
\be
\z =  -2{\epsilon}  \sum_{a=0}^n \hat \a_a \ln { P(z_a)} \,.
\ee


\section{Explicit evaluation of the partition function $Z_{(0:n)}$} 
\label{partition}
One may find the explicit form of the partition function using 
\eqref{maineq}.  
In this section we present how to obtain the partition function in a systematic way. 

To obtain $Z_{(0:n)}$, we compare each coefficient of order $z^l$ in \eqref{maineq}.
For the power $z^{N-2-k}$ with $0\leq k\leq N+n-1$, one has 
\be
    P_N d_k+ P_{N-1}d_{k-1} +\dots+ P_{N-k}d_{0}
=  {\epsilon}^2 \,(N-k) (N-k-1)P_{N-k}+2{\epsilon}\sum_{l=0}^k
( \hat c_{k-l}(N-l)P_{N-l} )\,.
\label{Pk}
\ee
Here we use the notation that $P_{a}$ vanishes when $a<0$ or $a>N$.
The highest power  $ z^{N-2}$ shows that
\be 
   d_0 =  {\epsilon}^2 \,N(N-1) +2{\epsilon} \hat c_0N \,,
\ee
which is independent of $\hat c_{k>0}$.
Finding $d_{k>0}$ needs algebraic manipulation. 

We  present the case  rank 2 ($n=2$) explicitly, which has $d_1$ only.
First note that the partition function is given in terms of differential equation \eqref{fd}, 
 \begin{align}
- g_s^2(\hat c_1\frac{\partial}{\partial \hat c_1} + 2\hat c_2 \frac{\partial}{\partial \hat c_2}) \log Z_{(0:2)}  &= d_0 \,,
\label{eqv0} \\
- g_s^2\hat c_2 \frac{\partial}{\partial \hat c_1} \log Z_{(0:2)} &= d_1\,. 
\label{eqv1}
\end{align}
Eq.~\eqref{eqv0} is solved to get
\be
g_s^2 \log Z_{(0:2)} = -\frac{d_0}{2}  \log \hat c_2 + h(\t ) \,,
\label{tau}
\ee 
where $h(\t)$ is any function of $\t \equiv \hat c_2/\hat c_1^2 $
which satisfies automatically  $v_0 ( h(\t))  =0$.
It is noted that the right hand side of eq.~\eqref{tau} is equivalent to $H_{(0:2)}$
given in \eqref{0nh}
up to normalization.
Eq.~\eqref{eqv1} requires $h(\t)$ to satisfy  
\be
d_1 =2g_s^2 ~ \hat c_1 \t^2 \frac{\partial \log Z } {\partial \t} 
= 2\hat c_1 \t^2 h'(\t)  \,.
\ee
This hints  that $\tilde d_1 =d_1/\hat  c_1$ should be a function of $\t$ only and one has 
$ h'(\t) = \frac{\tilde d_1 }{2  \t^2}  $ 
which can be solved as 
\be 
h(\t) = \frac1 {2  } \int^{\t  } d\t ~ {\tilde d_1 }/{\t^2}  \,.
\label{H}
\ee
Therefore it is enough to find $\tilde d_1 $ as a function $\tau$.  As described in appendix \ref{Method 1} we find
\begin{align}
\tilde d_1
& = 2 \epsilon   N + \t a_N  + \t^2  \frac{a_N (a_N -a_{N-1})}{2 \epsilon  }
\nn\\
&~~~
 +\t^3  \frac{ (a_N -a_{N-1})^2  -a_{N-1}(a_N -a_{N-1}) /2 + a_N (a_N -a_{N-1} )}
{(2 \epsilon  )^2 }
+{\cal O} ( \t^4)\,.
\end{align}
Once $\tilde d_1 $ is known, one can put $h(\t)$ in \eqref{H}as 
\be
h(\t) = \frac1{2  }
\left( -\frac{\tilde d_1^{(0)}}\t +  \tilde d_1^{(1)}\ln \t 
+ \sum_{\ell\ge 2}\frac{ \tilde d_1^{(\ell)} }{\ell-1 }  ~\t^{\ell-1} \right)\,,
\ee
where we neglect the $\t$-independent term which will be absorbed into the normalization $N_{(0:2)}$. 
This provides the explicit partition function of rank 2:
\begin{eqnarray}
 Z_{(0:2)} &= N_{(0:2)}
(\hat c_2)^{- \frac{{\epsilon_1}^2 \,N(N-1) +2{\epsilon_1}\hat c_0N}{2g_s^2}} 
\left( \frac{\hat c_2}{\hat c_1^2}  \right) ^{\frac{-{\epsilon_1}N\left({\epsilon_1}(N-1)+\hat c_0 \right)}{g_s^2}   
} ~e^{-  \frac {\epsilon_1 N_{2} \hat c_1^2  } {g_s^2\hat c_2 }  + \frac{1}{g_s^2}\cO \left(\frac{\hat c_2}{\hat c_1^2}\right)   }.
\end{eqnarray}
$N_{(0:2)}$ is the normalization  factor  independent of $\t$.
This procedure demonstrates that finding $Z_{(0:n)}$ with $n>2$ is straight-forward.
On the other hand, it is to be noted that $Z_{(0:n)}$ has no filling fraction except $N$. 
This shows that $Z_{(0:n)}$ provides the solution of the one-cut case.
In addition, it will be nice to find $P(z)$ and $d_k$ in a more compact form.

\section{Explicit evaluation of the partition function $Z_{(m:n)}$} 
\label{con-block}
In this section  we evaluate  $Z_{(m:n)}$. Its potential derivative is given as 
$ V'(z) =  \sum_{k=-m}^{n} \frac { \hat c_k } {z^{k+1}} $ and 
therefore, $ f(z)=\sum_k {d_k} /{z^{2+k}}$ where $k$ runs from $-m$ to $n-1$.   

Power expansion of Eq. ~\eqref{maineq} 
provides $N+m+n$  equations corresponding to $N+m+n$  variables. 
Explicitly,  for the power of $ z^{N-k-2}$ with $-m\leq k \leq N+n-1$ one has 
the algebraic equation
\be 
2{\epsilon}
(N-k+n)\hat c_{n}P_{N-k+n}+ \sum_{s=-m}^{n-1}
\big((  2{\epsilon}(N-k+s)\hat c_{s}-d_s)P_{N-k+s}\big)+  {\epsilon}^2 \,(N-k) (N-k-1) P_{N-k}=0\,.
\ee
We use the same convention in the previous section: 
 $P_{N}=1$ and $P_{a}$ vanishes when $a<0$ or $a>N$. 
In addition, the coefficient $\hat c_\ell=0$ when $\ell \ge n+1$ or $\ell < -m$. 

One has the simple relation for  the highest power $ z^{N+m -2}$ ($k=-m$) 
\be
 d_{-m}=2{\epsilon} N\hat c_{-m}  \label{gnm2}\,,
\ee
and for the lowest power $z^{-(n+1)}$  ($k=N+n-1$) 
\be
 P_{0}d_{n-1}= 2{\epsilon} \hat c_{n}P_{1}\label{gn1}.
\ee

Let us consider the case  $(m,n)=(2,2)$ for concreteness. 
In this situation we need $d_{-1},d_0$ and $d_1$ to find the partition function 
$Z_{(2:2)}$. 
The flow equations in \eqref{uv_k} read
\begin{align}
- g_s^2\hat c_{-2} \frac{\partial}{\partial \hat c_{-1}} \log Z_{(2:2)} 
&=d_{-1}  -2 \epsilon N \hat c_{-1} \,,\label{220}
\\
- g_s^2(\hat c_1\frac{\partial}{\partial \hat c_1}
 + 2\hat c_2 \frac{\partial}{\partial \hat c_2})\log Z_{(2:2)}  
&=  d_{0} \,,
\label{221}
\\
- g_s^2\hat c_2 \frac{\partial}{\partial \hat c_1} \log Z_{(2:2)}
&= d_{1}\label{222} \,.
\end{align}
Appendix \ref{Method 2} shows that 
\begin{align}
d_{-1} & =  2{\epsilon}\hat c_{-1}N -2{\epsilon}N\hat c_{-2} \eta + \cO (\y^2) \,,\\
d_{0}&=  2{\epsilon}\hat c_0N +  {\epsilon}^2 \,N(N-1) -2{\epsilon}N\hat c_{-1} \eta+ \cO  (\y^2) \,,\\
d_{1}&= 2{\epsilon}\hat c_1N -2{\epsilon}N\bigg({\epsilon_1}(N-1)
+\hat c_0\bigg)\y
\\
&\quad +2\bigg[{\epsilon}\hat c_{-1}N\y^2-\frac{\epsilon}{\hat c_1}N
\bigg({\epsilon}(N-1)+\hat c_0\bigg)\bigg(3{\epsilon}(N-1)
+2\hat c_0\bigg)\bigg]\y^2
+ \cO (\y^3)\,,
\end{align}
where $ \eta =\hat c_2/\hat c_1$.
Using the linear combination \eqref{221} $-\frac{1}{\y}\times$\eqref{222}, we have
\begin{align}\label{223}
- g_s^2( \hat c_2 \frac{\partial}{\partial \hat c_2})\log Z_{(2:2)}  &= 2{\epsilon}\hat c_0N +  \frac{3}{2}{\epsilon}^2 \,N(N-1) -{\epsilon}N\frac{\hat c_1}{\y}\\
-\bigg[{\epsilon}\hat c_{-1}N&+\bigg({\epsilon}(N-1)+\hat c_0\bigg)\bigg(3{\epsilon}(N-1)+2\hat c_0\bigg)\frac{{\epsilon}N}{2\hat c_1}\bigg]\y
+   \cO \big(\y^2\big)\nn\,.
\end{align}
Since  \eqref{220}, \eqref{222} and \eqref{223} are just simple derivative equations for $ \hat c_{-1}$, $ \hat c_{1}$ and $ \hat c_{2}$, we can easily find 
$H_{(2:2)}$ given in \eqref{0nh}.
\begin{align}
\label{logz22}
 H_{(2:2)}=
&-2{\epsilon}N\big({\epsilon}(N-1)+\hat c_0\big)\log \hat c_{1}
+\big(2{\epsilon}\hat c_0N +  \frac{3}{2}{\epsilon}^2 \,N(N-1) \big)\log \hat c_{2}+{\epsilon}N \frac{\hat c_1^2} {\hat c_2}\nn \\
& - \Big(2{\epsilon}N\hat c_{-1}
-\frac{\epsilon}{2\hat c_1}N\big({\epsilon}(N-1)+\hat c_0\big)\big(3{\epsilon}(N-1)+2\hat c_0\big) \Big) \frac{\hat c_2}{\hat c_1}
+   \cO \big(\y^2\big)\,.
\end{align}
Thus, one has the partition function
\begin{align}
& Z_{(2:2)} = N_{(2:2)}\times (\hat c_1)^{\frac{2{\epsilon}N\big({\epsilon}(N-1)+\hat c_0\big)}{g_s^2}} 
(\hat c_2)^{- \frac{2{\epsilon}\hat c_0N +  \frac{3}{2}{\epsilon}^2 \,N(N-1)}{g_s^2}}
\nn\\ 
&~~~~~~\times e^{ \frac1{g_s^2}
\big\{{2{\epsilon}N\hat c_{-1}\frac{\hat c_2}{\hat c_1}
-{\epsilon}N\frac{\hat c_1^2} {\hat c_2} -\frac{\epsilon}{2}N\big({\epsilon}(N-1)+\hat c_0\big)\big(3{\epsilon}(N-1)+2\hat c_0\big)\frac{\hat c_2}{\hat c_1^2} +\cO \big(\y^2\big)  }\big \}}\,.
\end{align}
Here, $N_{(2:2)}$ is the normalization  factor\footnote{
The normalization factor can be a function of   $\hat c_{0}$  and  $\hat c_{-2}$ 
since their derivatives are not given by the flow equations. However, there is no evidence 
for this parameter to be divergent.}. $\cO \big(\y^k\big) $ are polynomials of $\y$, and they satisfy the following conditions:\\ 
1. They can be determined completely by the group of equations  \eqref{nk2} with given $N$, using the perturbation method;\\
2. They are independent of $g_s$, i.e., they are finite because all the coefficients in the  group of equations are independent of $g_s$. \\
In this way, \eqref{logz22} leads directly to the fact $
\lim\limits_{g_s \to 0} g_s^2 \log \left(  {Z_{(2:2)}}/ {N_{(2:2)}}  \right) \to finite $.

\section{Conclusion}  
\label{conclusion}
Using the second order differential equation \eqref{maineq} 
of the polynomial $P(z)$,  we find a straightforward method
 to calculate classical ICB, by assuming 
a hierarchical ordering in $\hat c_k$ 
so that $P_k$ can be treated perturbatively.
 Compare to known methods, this new approach is efficient 
since $ P(z) $ is a  polynomial with a finite degree $N$,
 which leads to a finite number of equations with exact solutions. 
This property allows us to give a rigorous proof 
for the classical behavior of conformal blocks. 
Besides, the classical limit for Nekrasov partition function
is proposed  in \cite{NS} as 
$
{\cal Z}_{Nek}
\;\stackrel{\e_2 \to 0}{\sim}\;
\exp\left\lbrace \frac{1}{\e_2} {\cal W}_{Nek}\right\rbrace
$, with ${\cal W}_{Nek}$ finite.
This is naturally equivalent to classical ICB's exponential behavior, 
through the connection of AGT conjecture. 
We also expect that similar discussions can be applied to W-symmetry in future.

\subsection*{Acknowledgements}
This work is supported by the National Research Foundation of Korea(NRF) grant funded by the Korea government(MSIP) (NRF-2014R1A2A2A01004951).

\appendix
\section{Normalization}
\label{Normalization}
In the text, we skip the normalization factor  $N_{(0:n)}$  
when  the conformal block ${\cal F}_\Delta^{(m:n)} $ is considered
since normalization does not contribute.
It is noted that  the duality $b\to 1/b$ holds for  ${\cal F}_\Delta^{(m:n)} $.
The duality is obvious since the loop equation \eqref{eq_loop}  
is manifestly dual ($Q$ is dual in $b \to 1/b$).

On the other hand, the partition function \eqref{eq_beta-matrix} does not look invariant. 
However, as  $g_s\to 0$ and $b \to \infty$ (NS limit),  the partition function allows the 
perturbation expansion, and 
the normalization can be taken with $\hat c_{k \neq 0} \to 0$,
\be
N_{(0:n)}(\hat c_0)  =     \int \left( \prod_{I=1}^N d \lambda_I \right)
 \prod_{I < J} (\lambda_I - \lambda_J)^{2\b} 
            e^{ \frac{\sqrt{\beta}}{g}  \sum_I \hat c_0 \log \lambda_I}\,.
 \ee
This is given in terms of the Selberg integral \cite{0710.3981}
\begin{align}
S_N(\alpha,\d,\b) & \equiv
\int_0^1 \cdots \int_0^1 \,
 \prod_{i=1}^N t_i^{\alpha-1}(1-t_i)^{\d-1}
\prod_{1\le i < j\le n}{\vert t_i - t_j\vert }^{2\b}\,
 t_1\cdots t_n \\
&\phantom{\equiv}=
\prod_{j=0}^{N-1} \frac{\Gamma (\alpha+j\b)
\Gamma(\d+j\b)\Gamma(1+(j+1)\b)}
{\Gamma(\alpha+\d+(N+j-1)\b)\Gamma(1+\b)}\,,
\nn
\end{align}
with $\alpha =1+ \sqrt{\beta}\hat  c_0/g$ and 
$\delta =1$, 
one has\footnote{There is some ambiguity in the integration range for IMM, which could be adjusted by rescaling $\lambda_I$, and does not effect our result shown here.}
\be
N_{(0:n)}(\hat c_0) 
=\prod_{j=0}^{N-1} \frac{\Gamma (1+A\b+j\b)
\Gamma(1+j\b)\Gamma(1+(j+1)\b)}
{\Gamma(2+A\b+(N+j-1)\b)\Gamma(1+\b)}\,,
\ee
where $A=2 \hat c_0 / \e$ .

Obversely $\lim\limits_{g_s \to 0} g_s^2 \log N_{(0:n)}$ is equivalent to $\lim\limits_{\b \to \infty} (\log N_{(0:n)}) /\b$.
 Then using the property of Gamma function $\log\Gamma (a+Z)=z \log( z )-z +\cO ( \log( z ))$ when $z$ is large and $a$ is small, 
and making use of
\ba
\sum_{j=0}^{N-1} \left\{(A+j)+j+(j+1)\right\}
 =  
\sum_{j=0}^{N-1} 
 \left\{(A+N+j-1)+1\right\}\,,
 \ea
we find 
\begin{align}
&\lim\limits_{\b \to \infty} \frac{ \log N_{(0:n)}(\hat c_0) }\b =  \nn\\
&\lim\limits_{\b \to \infty}\sum_{j=0}^{N-1} (A+j)\log (A\b+j\b)+j\log(j\b)+(j+1)\log((j+1)\b)\\
&-(A+N+j-1)\log(A\b+(N+j-1)\b)-\log(\b)\nn\\
 & =\sum_{j=0}^{N-1} (A+j)\log (A+j)+j\log(j)+(j+1)\log((j+1))\nn\\
&-(A+N+j-1)\log(A+(N+j-1))\nn\\
& =const.\nn
\end{align}
This means $\lim\limits_{g_s \to 0} g_s^2  \log  N_{(0:n)}(\hat c_0) =const.$ 

For the regular case,
$N_{\beta} $ is the normalization factor independent on $z_a$,
 which means it can be achieved by setting all the $z_a=0$. Actually
\ba
N_{\beta}&&=     \int \left( \prod_{I=1}^N d \lambda_I \right)
 \prod_{I < J} (\lambda_I - \lambda_J)^{2\b} 
            e^{ \frac{\sqrt{\beta}}{g}  \sum_I \sum_{a=0}^{n} \hat \a_a \log\lambda_I}\nn \\
&&=   \int \left( \prod_{I=1}^N d \lambda_I \right)
 \prod_{I < J} (\lambda_I - \lambda_J)^{2\b} 
            e^{ \frac{\sqrt{\beta}}{g}  \sum_I \hat c_0 \log \lambda_I} \\
&&=N_{(0:n)}(\hat c_0)  \,. \nn
 \ea
According to the previous discussion we know
 $\lim\limits_{g_s \to 0} g_s^2 \log N_{\beta} =const.$ 
\section{Method to obtain $ d_1$}
\label{Method 1}
The eq.~\eqref{Pk} for rank 2 can be written as
\be
(d_1 - 2 \epsilon \hat c_1 t) P_t = \tilde a_t P_{t-1} + 2 \epsilon \hat c_2 (t+1) P_{t+1}\,,
\label{d1-1}
\ee
where we put  $N-k+1=t$  and 
 $\tilde a_t =( \epsilon^2  (t-2) +2 \epsilon \hat c_0 )(t-1) -d_0$.
One can simplify this if one puts $r_t = P_t/P_{t-1}$,
\be
d_1 = 2 \epsilon \hat c_1 t + 2 \epsilon \hat c_2 (t+1) r_{t+1} + \tilde a_t /r_t\,,
\ee
where $ t$ runs from 0 to $N$. 
From the definition, one has $r_{N+1}=0$ and $r_0 = \infty$. 

When $t=0$, one finds 
$d_1=  2 \epsilon \hat c_2~ r_1 $ 
in a very compact notation. 
However, explicit form of $r_1$ as the function of $\hat c_1, \hat c_2$  is not easy to put.
One way to find $d_1$ is to use perturbation. 
One may rescale $ r_t = \hat c_1  (N+1-t)  \xi_t/(\hat c_2 t) $ in \eqref{d1-1}
to get
\be
 d_1/\hat c_1  = 2 \epsilon   ( t + (N-t) \xi_{t+1} ) + \t  ~ { a_t }/{\xi_t}\,,
\label{d1-2}
\ee
where $a_t =   \tilde a_t~ t /(N+1-t)$
and $\t=\hat c_2/\hat c_1^2$ as defined in \eqref{tau}.
Eq.~\eqref{d1-2}  shows that $d_1/\hat c_1$ is indeed 
a function of $\t$.

To the lowest order in $\t$,  $\xi_t =1 $ and $ d_1 /\hat c_1 =  2 \epsilon   N$.
Therefore, one may find   $  d_1/\hat c_1 $ and $\xi_t$  in powers of $\t$
\be
\tilde d_1 := d_1/\hat c_1 = \sum_{\ell\ge 0 } \tilde d_1^{(\ell )} \t^\ell\,,~~~
\xi_t  = \sum_{\ell\ge 0 } \xi_t^{(\ell)} \t^\ell\,,
\ee
where $\tilde d_1^{(0)}  =   2 \epsilon   N$
and $\xi_t^{(0)}=1$. 

In addition, the solution of $\tilde d_1$  is $t$-independent.
Therefore, the perturbative  expansion is more facilitated 
if the equation is set into  the form 
\be
\tilde d_1  =   \tilde d_1^{(0)} +  \t  a_N + 
\Big[ 2 \epsilon   (N-t) (\xi_{t+1} -1)  +  \tau  (a_t-a_N) \Big] 
+ \t  a_t (1 - \xi_t)/\xi_t  \,.
\ee
where decomposition 
$1/\xi_t = 1 + (1-\xi_t)/\xi_t$ is used to put $1/\xi_t $ perturbatively
in $t$.
To make $\tilde d_1$ $t$-independent, one has 
$\tilde d_1^{(1)}=   a_N$ and the term in the squared bracket need to vanish at the order $\eta$ ,
\be
 2 \epsilon \hat c_1 (N-t) \xi_{t+1}^{(1)} +a_t -a_N =0\,,
\ee
which fixes the $\xi_t^{(1)} $. 
In this way one can find $\tilde d_1 $ order by oder  
\begin{align}
\tilde d_1
& = 2 \epsilon   N + \t a_N  - \t^2  \frac{a_N (a_N -a_{N-1})}{2 \epsilon  }
\nn\\
&~~~
 +\t^3  a_N \frac{ (a_N -a_{N-1})^2  -a_{N-1}(a_N -a_{N-2}) /2 + a_N (a_N -a_{N-1} )}
{(2 \epsilon  )^2 }
+{\cal O} ( \t^4)\,,
\end{align}
which provides the explicit $\tilde d_1^{(\ell)}$ with $\ell=0,1,2,3$.
\section{Method to obtain $d_{-1},d_0$ and $d_1$}
\label{Method 2}
For the case  $(m,n)=(2,2)$, explicitly expanding  Eq. \eqref{maineq}   in powers of $z$, we have
 \ba
&&  z^{N}: \qquad   d_{-2}=2{\epsilon} N\hat c_{-2}\\
&&  z^{N-1}: \qquad  d_{-1}  =  2{\epsilon}N\hat c_{-1} -2{\epsilon}\hat c_{-2}P_{N-1}\\
&&  z^{N-2}: \; d_{0}  =  2{\epsilon}N\hat c_0 +  {\epsilon}^2 \,N(N-1)+ ( 2{\epsilon}(N-1)\hat c_{-1}-d_{-1})P_{N-1}-4{\epsilon}\hat c_{-2}P_{N-2}\\
&&  z^{N-3}: \qquad  d_{1}  =  2{\epsilon}\hat c_1N +\big(  2{\epsilon}(N-1)\hat c_{0}+  {\epsilon}^2 \,(N-1) (N-2) -d_0\big)P_{N-1}\\
&&\qquad\qquad\qquad+\big(  2{\epsilon}(N-2)\hat c_{-1}-d_{-1}\big)P_{N-2}-6{\epsilon}\hat c_{-2}P_{N-3}\,.
\nn
\ea
And in general for the power of $ z^{N-k-2}$ with $-1\leq k \leq N+1$ we have
\ba 
\label{nk2}
&&2{\epsilon}
 (N-k+2)\hat c_{2}P_{N-k+2}+\big(  2{\epsilon}(N-k+1)\hat c_{1}-d_1\big)P_{N-k+1}\nn\\
&& +\big(  2{\epsilon}(N-k)\hat c_{0}+  {\epsilon}^2 \,(N-k) (N-k-1) -d_0\big)P_{N-k}\\
&&+\big(  2{\epsilon}(N-k-1)\hat c_{-1}-d_{-1}\big)P_{N-k-1}-2{\epsilon}(k+2)\hat c_{-2}P_{N-k-2}=0\,.\nn
\ea

To find the $d_i$'s, one may use perturbation in  \eqref{nk2}. 
First note that $P_{N-1}=  -\Big\langle \sum_{ I} \lambda_I \Big\rangle$ while 
$ P_{0}=\Big\langle \prod_{ I} (- \lambda_I)\Big\rangle$. 
Therefore, $P_{N-k}$ can grow as   $k$-powers of the expectation values. 
One may assume $ P_{N-t}  = \cO (\y^{t})$ with
$\vert\y \vert \ll1$. 
Indeed, the equation \eqref{nk2}  has the solution to the lowest order, 
 \ba
&&  z^{N-1}: \qquad  d_{-1}  =  2{\epsilon}\hat c_{-1}N + \cO (\y)\\
&&  z^{N-2}: \qquad  d_{0}  =  2{\epsilon}\hat c_0N +  {\epsilon_1}^2 \,N(N-1)+ \cO (\y)\\
&&  z^{N-3}: \qquad  d_{1}  =  2{\epsilon}\hat c_1N + \cO (\y)\\
&&  z^{N-4}: \qquad P_{N-1}  =\frac{\hat c_2}{\hat c_1} N + \cO (\y^2)\\
&&  z^{N-5}: \qquad P_{N-2}  =\frac{N -1}{2}\frac{\hat c_2}{\hat c_1}P_{N-1}  + \cO (\y^{3})\\
&&  z^{N-t-3}: \qquad P_{N-t}  =\frac{N -t+1}{t}\frac{\hat c_2}{\hat c_1}P_{N-t+1}  + \cO (\y^{t+1}).
\ea
Obviously given the condition $\vert{\hat c_2}/{\hat c_1} \vert \ll1$, while keeping $c_1$, $c_0$, $c_{-1}$ and $c_{-2}$  in the same order,
we can choose $\y \equiv {\hat c_2}/{\hat c_1}$, 
consistent with $ P_{N-t}  = \cO (\y^{t})$.

Next order is given as follows:
 \ba
&&  z^{N-1}: \qquad  d_{-1}^{(1)} =  -2{\epsilon}\hat c_{-2}N 
\nn\\
&&  z^{N-2}: \qquad  d_{0}^{(1)} =  -2{\epsilon}\hat c_{-1}N 
\nn\\
&&z^{N-3}: \qquad   d_1^{(1)}=-2{\epsilon}N\bigg({\epsilon}(N-1)+\hat c_0\bigg)
\ea
To calculate the partition function up to $\cO  (\y)$, for a technical reason which will be clear by practical application, we need the $\y^2$ expansion of $d_{1}$, which could be obtained from the third order perturbation:
 \ba
&&z^{N-3}: \qquad   d_1^{(2)}=
2{\epsilon}\hat c_{-1}N-\frac{\epsilon}{\hat c_1}N\bigg({\epsilon}(N-1)+\hat c_0\bigg)\bigg(3{\epsilon}(N-1)+2\hat c_0\bigg)
\ea
so that 
\begin{align}
d_{-1} & =  2{\epsilon}\hat c_{-1}N -2{\epsilon}N\hat c_{-2} \eta + \cO (\y^2)\\
d_{0}&=  2{\epsilon}\hat c_0N +  {\epsilon}^2 \,N(N-1) -2{\epsilon}N\hat c_{-1} \eta+ \cO  (\y^2)\\
d_{1}&=2{\epsilon}\hat c_1N -2{\epsilon}N 
\bigg({\epsilon}(N-1)+\hat c_0\bigg) \eta +  d_1^{(2)}\y^2 +\cO  (\y^3)
\end{align}

\end{document}